\documentclass[11pt]{article}
\usepackage{array}
\usepackage{xcolor}
\usepackage{verbatim}
\usepackage{amssymb}
\usepackage{mathrsfs}
\usepackage{multirow}
\usepackage{amsmath}
\usepackage{mathenv}
\usepackage[english]{babel}
\usepackage{listings}
\usepackage{graphicx}
\usepackage{float}
\usepackage{subfig} 
\usepackage{caption}
\usepackage{authblk}
\usepackage{amsthm}
\usepackage[nottoc]{tocbibind}
\providecommand{\keywords}[1]
{
	\small	
	\textbf{\textit{Keywords---}} #1
}
\newtheorem{theorem}{Theorem}[section]

\newtheorem{proposition}[theorem]{Proposition}

\newtheorem{remarque}[theorem]{Remarque}
\begin{document}	
	
	\title{An optimal control study for a two-strain SEIR epidemic model with saturated incidence rates and treatment}
		
	\author{Karam Allali$^a$, Mouhamadou A.M.T. Balde$^b$ and Babacar M. Ndiaye$^b$}
	
	\affil{$^a$ Laboratory of Mathematics, Computer Science and Applications, Faculty of Sciences and Technologies, University Hassan II of Casablanca, PO BOX 146, Mohammedia, Morocco. 
		{\it karam.allali@fstm.ac.ma}\\
		$^b$ Laboratory of Mathematics of Decision and Numerical Analysis, University of Cheikh Anta Diop, Dakar, Senegal.
 {\it mouhamadoualioumountagatall.balde@ucad.edu.sn, babacarm.ndiaye@ucad.edu.sn}
	}
	\maketitle
	
\begin{abstract}
This work will study an optimal control problem describing the two-strain SEIR epidemic model. The studied model is in the form of six nonlinear differential equations illustrating the dynamics of the susceptibles and the exposed, the infected, and the recovered individuals. The exposed and the infected compartments are each divided into two sub-classes representing the first and the second strain.   The model includes two saturated rates and two treatments for each strain. We begin our study by showing the well-posedness of our problem. The basic reproduction number is calculated and depends mainly on the reproduction numbers of the first and second strains. The global stability of the disease-free equilibrium is fulfilled. The optimal control study is achieved by using the Pontryagin minimum principle. Numerical simulations have shown the importance of therapy in minimizing the infection's effect. By administrating suitable therapies, the disease's severity decreases considerably. The estimation of parameters as well as a comparison study with COVID-19 clinical data are fulfilled. It was shown that the mathematical model results fits well the clinical data.
\end{abstract}
	

\section{Introduction}
\begin{figure}[!h]
		\centering
		\includegraphics[width=11cm,height=8cm]{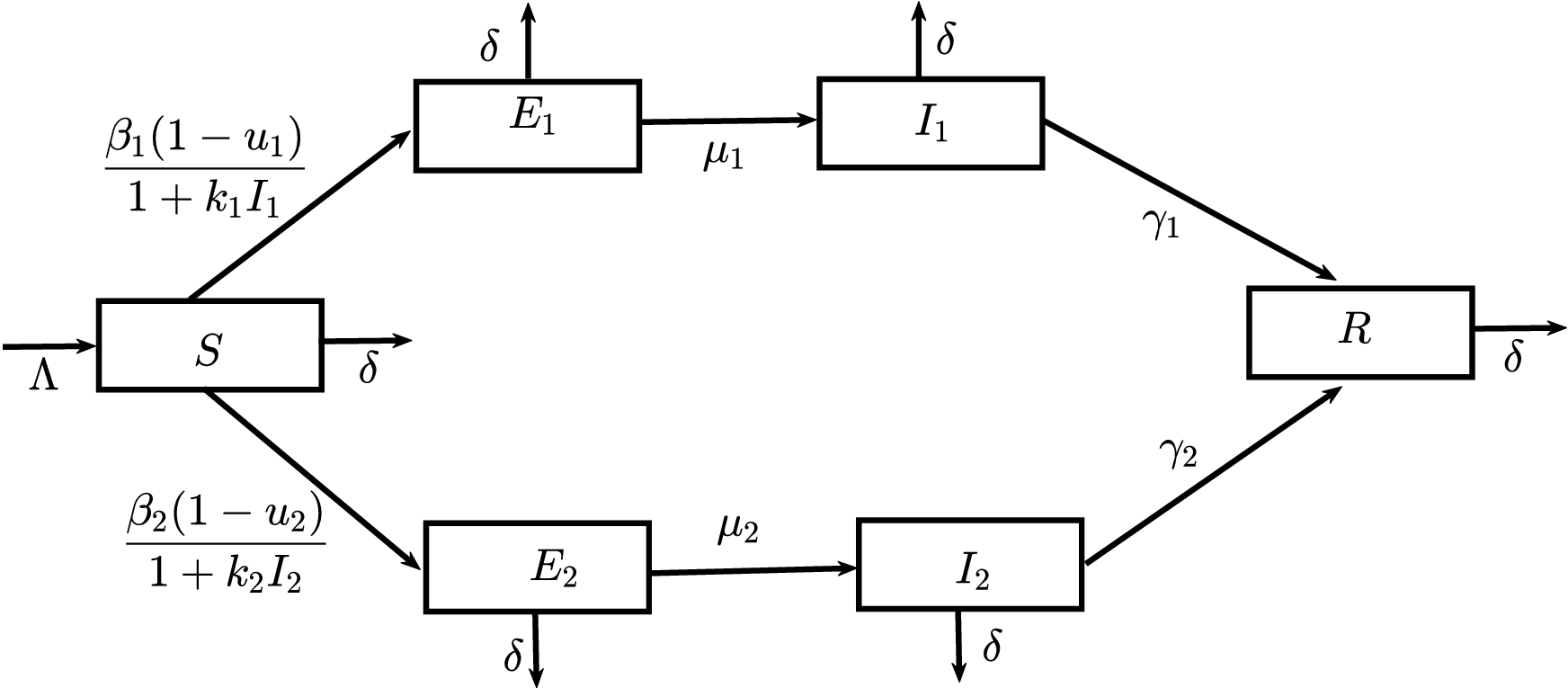}
		\caption{The two-strain infection diagram.}
		\label{fig1}
	\end{figure}

The SIR pioneer work of  Kermack and McKendric in 1927  \cite{kermack1927contribution} has inaugurated most mathematical modeling in epidemiology. SIR abbreviation stands for susceptible-infected-recovered compartments. Those classes were sufficient enough to study many infectious diseases. However, after the infection occurs, the individual is first exposed to the virus and will be infectious. Therefore, another compartment must be added to the classical SIR one, and the new abbreviation will be SEIR, which means susceptible-exposed-infected-recovered configuration. Many papers have used the SEIR model to study many infectious diseases such as coronavirus disease 2019 (COVID-19) \cite{he2020seir,dolbeault2020heterogeneous}, hepatitis B infection (HBV) \cite{side2017seir}, human immunodeficiency virus (HIV)\cite{guseva2023building} and many others.\\	
Mutation process was noticed in many infections such as influenza, dengue fever, COVID-19, HIV, and tuberculosis \cite{cov0,cov1,cov2,cov3,cov4,cov5,cov6,nuno2008mathematical,gubler1997dengue,coronaviridae2020species,brenchley2006microbial,golub2006delayed}. The mutation phenomenon is linked to the observation of the pathogen multiple strains. Consequently, mathematical models incorporating two or more strains are better for studying and understanding the evolution of many strains in a single disease case. The global dynamics of the multi-strain SEIR epidemic model with an application to the COVID-19 pandemic was studied in \cite{khyar2020global}. More recently, the analysis and optimal control of a two-strain SEIR infection model was studied in \cite{bentaleb2023analysis}. The authors of the latter work have considered that the infected individuals may recover at a saturated treatment and studied the optimal control of the strategies undertaken. In this work, we will continue the investigation of the optimal control of two-strain epidemic problems by considering two controls for minimizing the force of infection. In practice, the two controls can represent the susceptibles' vaccination or social distancing between the susceptible and the infectious individuals. The dynamics of our two-strain epidemic model with treatments and saturated rate are given under the following form:
	\begin{equation}\label{prob}
		\left \{
		\begin{array}{l c l}
			\displaystyle \frac{dS}{dt} &=& \Lambda - \displaystyle\frac{\beta_1 (1-u_1) S I_1}{1+k_1 I_1}  - \displaystyle\frac{\beta_2 (1-u_2) S I_2}{1+k_2 I_2} - \delta S, \\\\
			\displaystyle\frac{dE_1}{dt} &=& \displaystyle\frac{\beta_1 (1-u_1) SI_1}{1+k_1 I_1}-(\mu_1+\delta)E_1, \\\\
			\displaystyle\frac{dE_2}{dt} &=&  \displaystyle\frac{\beta_2 (1-u_2) SI_2}{1+k_2 I_2}-(\mu_2+\delta)E_2, \\\\
			\displaystyle\frac{dI_1}{dt}&=& \mu_1E_1-(\gamma_1 +\delta)I_1, \\\\
			\displaystyle\frac{dI_2}{dt}&=& \mu_2E_2-(\gamma_2 +\delta)I_2, \\\\
			\displaystyle\frac{dR}{dt} &=& \gamma_1 I_1 + \gamma_2 I_2 - \delta R. \\
		\end{array}
		\right.
	\end{equation}
Where the susceptible are denoted $S(t)$, the strain-1 exposed individuals by $E_1(t)$, the strain-2 exposed by $E_2(t)$, the strain-1 infected by $I_1(t)$, the strain-2 infected by $I_2(t)$ and the recovered individuals are represented by $R$. The parameter $\Lambda$ represents the recruitment rate of susceptibles through either birth or immigration, $\beta_1$ (respectively, $\beta_2$) is the strain-1 infection rate (respectively, the strain-2 infection rate). $\mu_1$  (respectively, $\mu_2$) is the strain-1 latency rate (respectively, the strain-2 latency rate). The parameter $k_i$ (i=1,2) measures the psychological, inhibitory, or crowding effect in the $i^{th}$ strain. $\gamma_2=i$ (i=1,2) is the $i^{th}$ strain recovery rate. The natural mortality rate of the population is denoted by $\delta$. Finally, the new parameters to the problem $u_1$ and $u_2$ represent the treatment efficiency for the first and the second strain, respectively. The two-strain diagram of the infection is illustrated in Fig. \ref{fig1}. The estimation of parameters as well as a comparison study with COVID-19 clinical data are fulfilled. It was shown that the mathematical model results fits well the clinical data.\\
	
\noindent The organization of this present paper will take the following form. First, Section 2 will be devoted to our two-strain SEIR epidemic model with two saturated incidence rates and two treatments. Next, Section 3 will investigate the disease-free equilibrium global stability. The optimal control study of our problem using the Pontryagin minimum principle is fulfilled in Section 4. Section 5 is dedicated to several numerical solutions showing the importance of the control strategy. The parameters estimation is given in Section 6. Concluding remarks are given in the section 7.

\section{The problem well-posedness}
In this section, we will focus attention on the problem's well-posedness. More precisely, we will show that the problem  \eqref{prob} admits a positive and bounded solution.
\subsection{Positivity result}
Our problem describes population dynamics; therefore, all the acting variables must be nonnegative. Hence, we have the following result of positive invariance.
\begin{theorem}
		For any initial data  $(S(0),E_1(0),E_2(0),I_1(0),I_2(0),R(0))\in \mathbb{R}^6_+$,  the variables $(S(t),E_1(t),E_2(t),I_1(t),I_2(t),R(t))$ to the model \eqref{prob} will remain also  positive $\forall \ \ t>0$.
	\end{theorem}
	
	\begin{proof}[\textbf{Proof}]
		First assume that $ t_f= sup\{\zeta \geq 0 \mid \forall t \in [0,zeta] \mbox{ we have } S(t)\geq 0, E_1(t)\geq 0, E_2(t)\geq 0, I_1(t)\geq 0, I_2(t)\geq 0 \mbox{ and } R(t)\geq 0 \}$. Let's prove now that $t_f= +\infty$.\\	
\noindent	Suppose that $0< t_f < +\infty$ then from the solutions continuity fact, we have: $S(t_f)=0$ or $E_1(t_f)=0$ or $E_2(t_f)=0$ or $I_1(t_f)=0$ or $I_2(t_f)=0$ or $R(t_f)=0$.\\
			Suppose that $S(t_f)=0$, then:
		
		\begin{equation}\label{posit}
			S(t_f)=0 \Rightarrow \displaystyle\frac{dS(t_f)}{dt}= \lim\limits_{t \rightarrow t_f^-} \frac{S(t_f)-S(t)}{t_f-t} = \lim\limits_{t \rightarrow t_f^-} \frac{-S(t)}{t_f-t} \leq 0
		\end{equation}
		However from the system \eqref{prob} first equation, we have $\frac{dS(t_f)}{dt}= \Lambda > 0$, which contradicts \eqref{posit}. Similar remark for the other variables $E_1(t), E_2(t), I_1(t), I_2(t)$ and $R(t)$. We conclude $t_f$ that could not be finite. This concludes the demonstration.
		
	\end{proof}
	
	\subsection{Boundedness result}
	
	\begin{proposition}
		The following closed-set
		\begin{eqnarray*}
			\mathcal{P} & = & \{(S(t),E_1(t),E_2(t),I_1(t),I_2(t),R(t)) \in \mathbb{R}^6_+, \mbox{ such that, }  \\
			& & S(t)+E_1(t)+E_2(t)+I_1(t)+I_2(t)+R(t) \leq\displaystyle\frac{\Lambda}{\delta} \}
		\end{eqnarray*}
		is a positively invariant set.
	\end{proposition}
	
	\begin{proof}[\textbf{Proof}]
		Let the total population
		$$
		T=S+E_1+E_2+I_1+I_2+R.
		$$
		Adding all equations of \eqref{prob} to each other, we will have:
		
		\begin{center}
			$\displaystyle\frac{dT}{dt} = \Lambda - \delta N, $
		\end{center}
		therefore  \begin{center}
			$T(t)=\displaystyle\frac{\Lambda}{\delta} + (T(0)-\displaystyle\frac{\Lambda}{\delta})e^{-\delta t} $
		\end{center}
		
\noindent	Hence, for $t \rightarrow \infty$, we will have $T(t)=\displaystyle\frac{\Lambda}{\delta}$.\\
\noindent	We conclude that $\mathcal{P}$ is a positively invariant set. All solutions to the problem \eqref{prob} remain bounded.	
	\end{proof}

\section{Analysis of the problem}
This section will be devoted to providing the basic reproduction number and the disease-free equilibrium stability result.\\
Since the five first equations of the problem \eqref{prob} are independent of the last variable $R$, we can omit the last equation. Hence, we can study the following five equations reduced system:
	\begin{equation}\label{prob2}
		\left \{
		\begin{array}{l c l}
			\displaystyle \frac{dS}{dt} &=& \Lambda - \displaystyle\frac{\beta_1 (1-u_1) S I_1}{1+k_1 I_1}  - \displaystyle\frac{\beta_2 (1-u_2) S I_2}{1+k_2 I_2} - \delta S, \\\\
			\displaystyle\frac{dE_1}{dt} &=& \displaystyle\frac{\beta_1 (1-u_1) SI_1}{1+k_1 I_1}-(\mu_1+\delta)E_1, \\\\
			\displaystyle\frac{dE_2}{dt} &=&  \displaystyle\frac{\beta_2 (1-u_2) SI_2}{1+k_2 I_2}-(\mu_2+\delta)E_2, \\\\
			\displaystyle\frac{dI_1}{dt}&=& \mu_1E_1-(\gamma_1 +\delta)I_1, \\\\
			\displaystyle\frac{dI_2}{dt}&=& \mu_2E_2-(\gamma_2 +\delta)I_2. \\
		\end{array}
		\right.
	\end{equation}
	
	\subsection{Basic reproduction number}
	First of all, it is clear that the system \eqref{prob2} has a unique disease-free equilibrium given as follows:
	
	\begin{equation}
		\mathcal{E}_{0} = (S^0,E_1^{0},E_2^{0},I_1^{0},I_2^{0}) = (\displaystyle\frac{\Lambda}{\delta}, 0, 0, 0, 0).
	\end{equation}
\noindent The next generation matrix method in \cite{van2002reproduction} will be used to calculate the problem \eqref{prob} basic reproduction number $\mathcal{R}_0$. As it is well known, we have:
	$$
	\mathbf{\mathcal{R}_0 = \rho (\mathcal{F}\mathcal{V}^{-1})}
	$$
	Where $\rho$ is the spectral radius, the matrix $\mathcal{F}$ describes the new infections, while the matrix $\mathcal{V}$ represents the transition terms.
\begin{center}
		$
		\mathbf{
			\mathcal{F}=
			\quad
			\begin{pmatrix}
				0&0&\beta_1 (1-u_1) S^0&0\\
				0&0&0&\beta_2 (1-u_2) S^0\\
				0&0&0&0\\
				0&0&0&0\\
			\end{pmatrix}
		}
		$
		,
		$
		\mathbf{
			\mathcal{V}=
			\quad
			\begin{pmatrix}
				\mu_1+\delta&0&0&0\\
				0&\mu_2+\delta&0&0\\
				-\mu_1&0&\gamma_1+\delta&0\\
				0&-\mu_2&0&\gamma_2+\delta\\
			\end{pmatrix}
		}
		$
	\end{center}
	So $\mathcal{F}\mathcal{V}^{-1}$ is given as :
	
	\begin{flushleft}
		$
		\mathcal{F}\mathcal{V}^{-1}=
		\quad
		\begin{pmatrix}
			\frac{\beta_1 (1-u_1) S^0 \mu_1}{(\mu_1+\delta)(\gamma_1+\delta)}&0&\frac{\beta_1 (1-u_1) S^0}{\gamma_1+\delta}&0\\
			0&\frac{\beta_2 (1-u_2) S^0\mu_2}{(\mu_2+\delta)(\gamma_2+\delta)}&0&\frac{\beta_2 (1-u_2) S^0}{\gamma_2+\delta}\\
			0&0&0&0\\
			0&0&0&0\\
		\end{pmatrix}.
		$
	\end{flushleft}
	Hence, the basic reproduction number is given by :
	\begin{equation}
		\mathcal{R}_0=max\{\mathcal{R}_{0,1},\mathcal{R}_{0,2}\},
	\end{equation}
	with,
	\begin{equation}
		\mathcal{R}_{0,1} = \frac{\Lambda \beta_1 (1-u_1) \mu_1}{\delta(\mu_1+\delta)(\gamma_1+\delta)}
		\quad \mbox{and} \quad
		\mathcal{R}_{0,2} = \frac{\Lambda \beta_2 (1-u_2) \mu_2}{\delta(\mu_2+\delta)(\gamma_2+\delta)}.
	\end{equation}
	$\mathcal{R}_{0,1}$ is called the strain-1 reproduction number, while $\mathcal{R}_{0,2}$ is nemad the strain-2 reproduction number.

	\subsection{Global stability of the disease-free equilibrium}
	We obtain the global stability of the disease-free equilibrium using the Lyapunov method. More precisely, we have the following result:
	\begin{theorem}
		The disease-free equilibrium $\mathcal{E}_0$  is globally asymptotically stable when the basic reproduction number is less than unity
	\end{theorem}

	\begin{proof}[\textbf{Proof}]
		First, Let's consider the following Lyapunov function:
		\begin{equation}
			L  = S-S^0 -S^0 ln(\frac{S}{S^0}) + E_1 + E_2 + \frac{\mu_1+\delta}{\mu_1}I_1 + \frac{\mu_2+\delta}{\mu_2}I_2.
		\end{equation}
		We remark that $L(\mathcal{E}_0)=0$. The positivity of  $L$ comes from the fact that
		
		\begin{center}
			$ ln \displaystyle\frac{S}{S^0} < \displaystyle\frac{S}{S^0}-1, \ \ \ \forall S \neq S^0.$
		\end{center}
		
		\begin{flushleft}
			Now, the time derivative of $L(t)$ is given by:
		\end{flushleft}
		
		\begin{equation}
			\dot{L} =(1-\displaystyle\frac{S^0}{S})\frac{d S}{dt} +\frac{d E_1}{dt}+\frac{d  E_2}{dt}+\frac{\mu_1+\delta}{\mu_1}\frac{d  I_1}{dt}+\frac{\mu_2+\delta}{\mu_2}\frac{d I_2}{dt}.
		\end{equation}
Hence
			\begin{flushleft}
			\begin{small}
				$
				\begin{array}{c l l}
					\dot{L} &=& (1-\displaystyle\frac{S^0}{S}) \left(\Lambda - \delta S- \sum \limits_{i=1}^2 \displaystyle\frac{\beta_i SI_i}{1+k_i I_i}\right)
					+\displaystyle\frac{\beta_1 (1-u_1) SI_1}{1+k_1 I_1} -(\mu_1+\delta)E_1\\
					&+& \displaystyle\frac{\beta_2 (1-u_2) SI_2}{1+k_2 I_2}-(\mu_2+\delta)E_2
					+(\mu_1+\delta)E_1  - \displaystyle\frac{(\gamma_1+\delta)(\mu_1+\delta)}{\mu_1} I_1\\
					&+&(\mu_2+\delta)E_2 - \displaystyle\frac{(\gamma_2+\delta)(\mu_2+\delta)}{\mu_2} I_2\\
					&&\\
					\dot{L} &=& \delta S^0(2-\displaystyle\frac{S}{S^0}-\displaystyle\frac{S^0}{S})
					+\displaystyle\frac{\beta_1 (1-u_1) S^0 I_1}{1+k_1 I_1}-\displaystyle\frac{(\gamma_1+\delta)(\mu_1+\delta)}{\mu_1} I_1 \\ 
					&+& \displaystyle\frac{\beta_2 (1-u_2) S^0 I_2}{1+k_2 I_2}-\displaystyle\frac{(\gamma_2+\delta)(\mu_2+\delta)}{\mu_2} I_2.
				\end{array}
				$
			\end{small}
		\end{flushleft}
		Hence, we have
		$$
		\dot{L} \le \delta S^0(2-\displaystyle\frac{S}{S^0}-\displaystyle\frac{S^0}{S})
		+\beta_1 (1-u_1) S^0 I_1-\displaystyle\frac{(\gamma_1+\delta)(\mu_1+\delta)}{\mu_1} I_1  + \beta_2 (1-u_2) S^0 I_2-\displaystyle\frac{(\gamma_2+\delta)(\mu_2+\delta)}{\mu_2} I_2.
		$$
		Which means that
		$$
		\dot{L} \le \delta S^0(2-\displaystyle\frac{S}{S^0}-\displaystyle\frac{S^0}{S})
		+ \displaystyle\frac{(\gamma_1+\delta)(\mu_1+\delta)I_1}{\mu_1}(\mathcal{R}_{0,1}-1)+\displaystyle\frac{(\gamma_2+\delta)(\mu_2+\delta)I_2}{\mu_2}(\mathcal{R}_{0,2}-1).
		$$
		Moreover, knowing that
		$$
		\displaystyle\frac{S}{S^0}+\displaystyle\frac{S^0}{S}\geq 2.
		$$
		It leads to
		$$
		\delta S^0(2-\displaystyle\frac{S}{S^0}-\displaystyle\frac{S^0}{S})\leq 0.
		$$
		Finally, $\dot{L} \le 0$ when $\mathcal{R}_{0,1} \le 1$ and $\mathcal{R}_{0,0} \le 1$.\\	
	We conclude that the negativity of the Lyapunov function will be achieved when the basic reproduction number is less than unity.\\
				The disease-free equilibrium is globally asymptotically stable when $\mathcal{R}_0\le 0$.
	\end{proof}
	
	\section{The optimal control study}
	To study the optimal control, we will vary the two treatments $u_1$ and $u_2$ and seek the optimal pair $(u^*_1,u^*_2)$ to reduce the infection. The model with the varied controls becomes:
	\begin{equation}\label{prob3}
		\left \{
		\begin{array}{l c l}
			\displaystyle \frac{dS}{dt} &=& \Lambda - \displaystyle\frac{\beta_1 (1-u_1(t)) S I_1}{1+k_1 I_1}  - \displaystyle\frac{\beta_2 (1-u_2(t)) S I_2}{1+k_2 I_2} - \delta S, \\\\
			\displaystyle\frac{dE_1}{dt} &=& \displaystyle\frac{\beta_1 (1-u_1(t)) SI_1}{1+k_1 I_1}-(\mu_1+\delta)E_1, \\\\
			\displaystyle\frac{dE_2}{dt} &=&  \displaystyle\frac{\beta_2 (1-u_2(t)) SI_2}{1+k_2 I_2}-(\mu_2+\delta)E_2, \\\\
			\displaystyle\frac{dI_1}{dt}&=& \mu_1E_1-(\gamma_1 +\delta)I_1, \\\\
			\displaystyle\frac{dI_2}{dt}&=& \mu_2E_2-(\gamma_2 +\delta)I_2 \\\\
			\displaystyle\frac{dR}{dt} &=& \gamma_1 I_1 + \gamma_2 I_2 - \delta R, \\
		\end{array}
		\right.
	\end{equation}
	where $u_i(t) \in [0,1], \ \ \forall t\ge 0$.
	\subsection{The infection optimization problem}
	Let the objective functional be maximized as follows:
	\begin{equation}
		J(u_1,u_2)= \int_{0}^{t_e}(S-(\frac{C_1}{2}u_1^2(t)+\frac{C_2}{2}u_2^2(t))dt.
	\end{equation}
	Where $t_e$ represents the needed end-time for the treatment measures. The two positive constants $C_1$ and $C_2$
	are based on the costs of each treatment strategy $u_{1}(t)$ and $u_{2}(t)$, respectively.\\
	\noindent The principal objective is to find the right pair $(u^*_1,u^*_2)$  that allows reducing the infection severity. In other words, we will seek to maximize the susceptible and to minimize the costs.
	
	\begin{equation}
		J(u^*_1,u^*_2)=\underset{\{u_1,u_2\} \in U}{max} J(u_1,u_2).
	\end{equation}
	Where $U$ is the control set given by:
	$$
	U=\{u_i(t)\mbox{ measurable}, \ \  0 \leq u_i \leq 1\mbox{ for }t \in [0,t_e], i=1,2\}.
	$$
	\subsection{Existence of trajectories}
	Let's set $x=(S(t),E_1(t),E_2(t),I_1(t),I_2(t),R(t))$ and let us denote by $f(t,x,u)=(f_{i})_{i=1,\cdots,6}$ the right hand side of the differential equation model \eqref{prob3}, with
	\begin{eqnarray*}
		f_1 &=& \Lambda - \displaystyle\frac{\beta_1 (1-u_1(t)) S I_1}{1+k_1 I_1}  - \displaystyle\frac{\beta_2 (1-u_2(t)) S I_2}{1+k_2 I_2} - \delta S, \\
		f_2 &=& \displaystyle\frac{\beta_1 (1-u_1(t)) SI_1}{1+k_1 I_1}-(\mu_1+\delta)E_1, \\
		f_3 &=& \displaystyle\frac{\beta_2 (1-u_2(t)) SI_2}{1+k_2 I_2}-(\mu_2+\delta)E_2, \\
		f_4 &=& \mu_1E_1-(\gamma_1 +\delta)I_1, \\
		f_5 &=& \mu_2E_2-(\gamma_2 +\delta)I_2, \\
		f_6 &=& \gamma_1 I_1 + \gamma_2 I_2 - \delta R.
	\end{eqnarray*}
	\begin{theorem}
		In the model \eqref{prob3}, there exists positive constants $K_1$ and $K_2$ such that $f$ satisfies :
		\begin{align}
			\left\{\begin{array}{l}
				\|f(t,x,u)\|\leq K_1(1+\|x\|+\|u\|) \\
				\|f(t,\tilde{x},u)-f(t,x,u)\|\leq K_2\|\tilde{x}-x\|(1+\|u\|)\\ 
			\end{array}
			\right.
			\label{lipsch1}
		\end{align} 
		for all $t\in [0,t_e]$, $\tilde{x},\ x\in \mathcal{P}$ and $u\in U$. 
		\label{edolipsch}
	\end{theorem}
	\begin{remarque}
		Since $x$ and $u$ belong in bounded set, we can replace \eqref{lipsch1} by 
		\begin{align}
			\left\{\begin{array}{l}
				\|f(t,x,u)\|\leq K \\
				\|f(t,\tilde{x},u)-f(t,x,u)\|\leq K\|\tilde{x}-x\|\\ 
			\end{array}
			\right.
			\label{lipsch2}
		\end{align}
		Morever in \eqref{prob3}, $f(t,x,u)$ is at least $C^1$ with respect to $t,\ x, \ u$ then \eqref{lipsch2} is satified see Fleming (\cite{Fleming}).  
	\end{remarque}
	\subsection{The infection optimality system}
	To apply Pontryagin’s minimum principle \cite{pontryagin1962mathematical}, we will need the following Hamiltonian under the following form
	$$
	{\cal H} = -S+(\frac{C_1}{2}u_1^2(t)+\frac{C_2}{2}u_2^2(t))+\sum_{i=1}^{6} \lambda_i f_i.
	$$
	The optimality system to the problem \eqref{prob3} is given by the following result
	\begin{theorem}
		There exist six adjoint equations to the problem \eqref{prob3} given by:
		\begin{equation}
			\begin{array}{c l l}
				\displaystyle\frac{d\lambda_1(t)}{dt}&=&1+\delta\lambda_1(t)+(\lambda_1(t)-\lambda_2(t))\displaystyle\frac{\beta_1(1-u_1)I_1}{1+k_1 I_1}+(\lambda_1(t)-\lambda_3(t))\displaystyle\frac{\beta_2(1-u_2) I_2}{1+k_2 I_2},\\
				\displaystyle\frac{d\lambda_2(t)}{dt}&=&(\lambda_2(t)-\lambda_4(t))\mu_1+\delta\lambda_2(t),\\
				\displaystyle\frac{d\lambda_3(t)}{dt}&=&(\lambda_3(t)-\lambda_5(t))\mu_2+\delta\lambda_3(t),\\
				\displaystyle\frac{d\lambda_4(t)}{dt}&=&(\lambda_1(t)-\lambda_2(t))\displaystyle\frac{\beta_1(1-u_1)S}{(1+k_1 I_1)^2}+(\lambda_4(t)-\lambda_6(t))\gamma_1+\delta\lambda_4(t),\\
				\displaystyle\frac{d\lambda_5(t)}{dt}&=&(\lambda_1(t)-\lambda_3(t))\displaystyle\frac{\beta_2(1-u_2)S}{(1+k_2 I_2)^2}+(\lambda_5(t)-\lambda_6(t))\gamma_2+\delta\lambda_5(t),\\
				\displaystyle\frac{d\lambda_6(t)}{dt}&=&\delta\lambda_6(t).
			\end{array}
		\end{equation}
		with the transversality conditions:
		\begin{center}
			$\lambda_i(t_e)=0$ for $i =1,...,6$.
		\end{center}
		The optimal controls are given by:
		\begin{eqnarray}
			u_1^* &=& min(1,max(0,\frac{1}{C_1}\displaystyle\frac{\beta_1 S I_1}{1+k_1 I_1} (\lambda_2(t)-\lambda_1(t)))), \\
			u_2^* &=& min(1,max(0,\frac{1}{C_2}\displaystyle\frac{\beta_2 S I_2}{1+k_2 I_2} (\lambda_3(t)-\lambda_1(t)))).
		\end{eqnarray}
	\end{theorem}
	\begin{proof}
		The six adjoint equations can be obtained via Pontryagin
		principle \cite{pontryagin1962mathematical}, such that
		\begin{equation}\label{adj}
			\begin{array}{c l }
				\displaystyle\frac{d\lambda_1(t)}{dt}=-\displaystyle\frac{\partial {\cal H}}{\partial S},\\
				\displaystyle\frac{d\lambda_2(t)}{dt}=-\displaystyle\frac{\partial {\cal H}}{\partial E_1},\\
				\displaystyle\frac{d\lambda_3(t)}{dt}=-\displaystyle\frac{\partial {\cal H}}{\partial E_2},\\
				\displaystyle\frac{d\lambda_4(t)}{dt}=-\displaystyle\frac{\partial {\cal H}}{\partial I_1},\\
				\displaystyle\frac{d\lambda_5(t)}{dt}=-\displaystyle\frac{\partial {\cal H}}{\partial I_2},\\
				\displaystyle\frac{d\lambda_6(t)}{dt}=-\displaystyle\frac{\partial {\cal H}}{\partial R},
			\end{array}
		\end{equation}
		with the transversality conditions:
		\begin{center}
			$\lambda_i(t_e)=0$ for $i =1,...,6$.
		\end{center}
		The system \eqref{adj} becomes
		\begin{equation}
			\begin{array}{c l l}
				\displaystyle\frac{d\lambda_1(t)}{dt}&=&1+\delta\lambda_1(t)+(\lambda_1(t)-\lambda_2(t))\displaystyle\frac{\beta_1(1-u_1(t))I_1}{1+k_1 I_1}+(\lambda_1(t)-\lambda_3(t))\displaystyle\frac{\beta_2(1-u_2(t)) I_2}{1+k_2 I_2},\\
				\displaystyle\frac{d\lambda_2(t)}{dt}&=&(\lambda_2(t)-\lambda_4(t))\mu_1+\delta\lambda_2(t),\\
				\displaystyle\frac{d\lambda_3(t)}{dt}&=&(\lambda_3(t)-\lambda_5(t))\mu_2+\delta\lambda_3(t),\\
				\displaystyle\frac{d\lambda_4(t)}{dt}&=&(\lambda_1(t)-\lambda_2(t))\displaystyle\frac{\beta_1(1-u_1)S}{(1+k_1 I_1)^2}+(\lambda_4(t)-\lambda_6(t))\gamma_1+\delta\lambda_4(t),\\
				\displaystyle\frac{d\lambda_5(t)}{dt}&=&(\lambda_1(t)-\lambda_3(t))\displaystyle\frac{\beta_2(1-u_2)S}{(1+k_2 I_2)^2}+(\lambda_5(t)-\lambda_6(t))\gamma_2+\delta\lambda_5(t),\\
				\displaystyle\frac{d\lambda_6(t)}{dt}&=&\delta\lambda_6(t).
			\end{array}
		\end{equation}
		under the transversality conditions:
		\begin{center}
			$\lambda_i(t_e)=0$ for $i =1,...,6$.
		\end{center}
		The two optimal controls can be found by solving:
		\begin{eqnarray}
			\displaystyle\frac{\partial {\cal H}}{\partial u_1} &=& 0, \\
			\displaystyle\frac{\partial {\cal H}}{\partial u_2} &=& 0.
		\end{eqnarray}
		This leads to
		\begin{eqnarray}
			C_1 u_1(t) +  \displaystyle\frac{\beta_1 S I_1}{1+k_1 I_1} (\lambda_1(t)-\lambda_2(t)) &=& 0, \\
			C_2 u_2(t) +  \displaystyle\frac{\beta_2 S I_2}{1+k_2 I_2} (\lambda_1(t)-\lambda_3(t)) &=& 0.
		\end{eqnarray}
		Since the two controls are in the interval $[0,1]$. Therefore
		\begin{eqnarray}
			u_1^* &=& min(1,max(0,\frac{1}{C_1}\displaystyle\frac{\beta_1 S I_1}{1+k_1 I_1} (\lambda_2(t)-\lambda_1(t)))), \\
			u_2^* &=& min(1,max(0,\frac{1}{C_2}\displaystyle\frac{\beta_2 S I_2}{1+k_2 I_2} (\lambda_3(t)-\lambda_1(t)))).
		\end{eqnarray}
	\end{proof}
	\section{Numerical Simulations}
	In this section, we will provide some numerical simulations to highlight the role of the control strategies in combating the infection. We use Gekko Optimization Suite \cite{gekko} in python \cite{Pyth}. 
	Fitting data of the cumulative cases of Senegal country estimates the model's parameters. For more detail, see the subsection \ref{estima}. We consider $T=16743927$ the total population of Senegal. We make two tests: test 1 and test 2. The values of the parameters are shown in the table :
	
	\begin{table}[H]
		\begin{center}
			\begin{tabular}{|c|c|c|c|}
				\hline 
				Parameters & test 1               & test 2  &       \\ 
				\hline                                                 
				$\Lambda $  & $0.0914 T/100$      & $ 0.0914 T/100$ & fixed                       \\ 
				\hline                                                                                       
				$k_{1} $  & $ 10$                   & $ 0.2$        & fixed              \\ 
				\hline                                                 
				
				$k_{2} $  & $ 5$                   & $ 0.15$        & fixed               \\ 
				\hline                            
				$\delta $  & $ 0.000219$                   & $ 0.000219$       & fixed                \\ 
				\hline                                                       
				$\xi_{1} $  & $ 0.08$                   & $ 0.08$        & fixed               \\ 
				\hline
				$\xi_{2} $  & $ 0.9$                   & $ 0.9$         & fixed              \\ 
				\hline 
				$\mu_{1} $  & $ 0.144$                   & $ 0.003$      & fitted                 \\ 
				\hline                                 
				$\mu_{2} $  & $ 0.196$               & $ 4.896\cdot 10^{-5}$         & fitted              \\ 
				\hline                                                         
				$\beta_1 $ & $ 3.68\cdot 10^{-7}$  & $ 1.52\cdot 10^{-6}$ & fitted \\ 
				\hline                                                         
				$\beta_2 $ & $ 2.25\cdot 10^{-8}$  & $ 4.6\cdot 10^{-6}$ & fitted \\ 
				\hline                                                         
				$\gamma_{1} $    & $ 0.19$        & $ 0.0009$    & fitted  \\ 
				\hline                                                         
				$\gamma_{2} $   & $ 0.17$           & $ 0.0002$   & fitted          \\ 
				\hline                            
							$C_{1} $   & $ 100/\beta_{2}$              & $ 100/\beta_{2}$   & fixed          \\ 
							\hline                                                 
							$C_{2} $   & $ 100/\beta_{1}$              & $ 100/\beta_{1}$   & fixed          \\ 
							\hline                                                                
			\end{tabular}
			\caption{Table of parameters: values of parameters for different tests.}
			\label{param}
		\end{center}
	\end{table}
	
	\noindent The following table gives initial conditions :
	\begin{table}[H]
		
		\begin{tabular}{|l|l|l|l|l|l|l|}
			\hline
			Initial conditions& $S(0)$&$E_{1}(0) $& $E_{2}(0) $&$I_{1}(0)$& $I_{2}(0)$&$R(0)$  \\
			\hline
			Test 1 & $16743677$&$172$&$10$&$34$&$34$&$0$ \\
			\hline
			Test 2 & $16661290$&$8085$&$40424$&$6826$&$27303$&$0$ \\
			\hline
		\end{tabular}\\
		\caption{Initial conditions}
	\end{table}

	\begin{figure}[H]
		\centering
		\subfloat[Susceptible $S$ and recovered $R$ cases.\label{fig:SR-EDO}]{\includegraphics[width=0.4\linewidth]{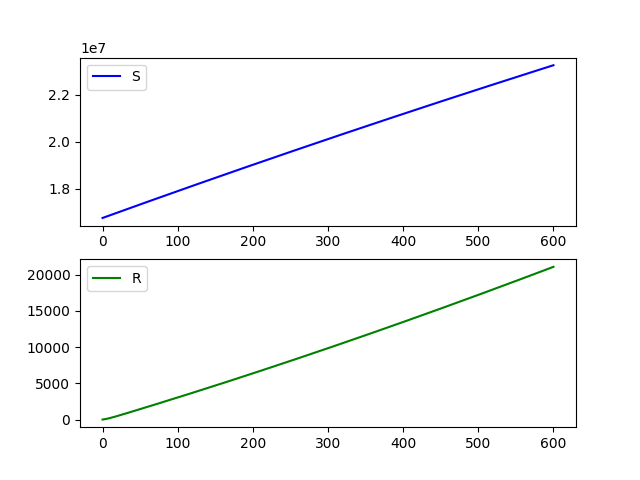}}\qquad\qquad 
		\subfloat[Susceptible $S$ and recovered $R$ cases.\label{fig:SR-Optcont}]{\includegraphics[width=0.4\linewidth]{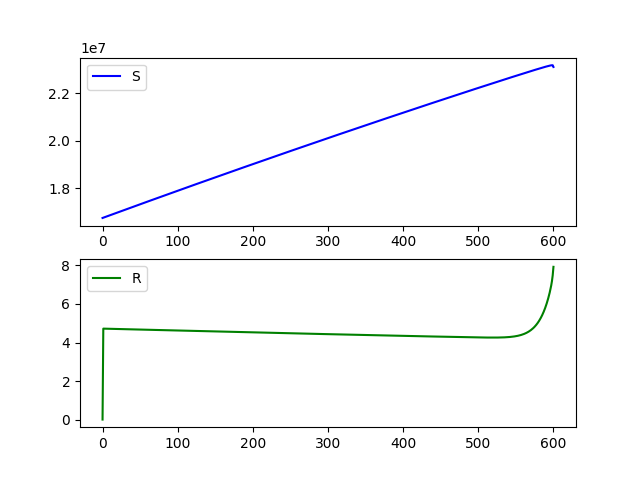}}\qquad\qquad \\
		\quad\subfloat[Exposed $E_{1}$ and $E_{2}$.\label{fig:E1E2-EDO}]{\includegraphics[width=0.4\linewidth]{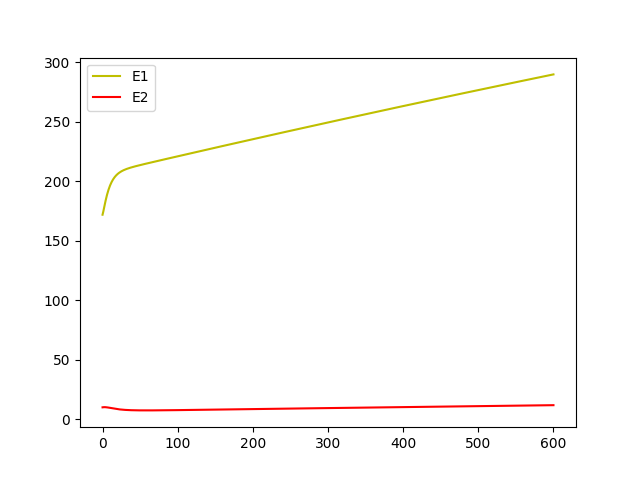}}\qquad\qquad
		\quad\subfloat[Exposed $E_{1}$ and $E_{2}$.\label{fig:E1E2-Optcont}]{\includegraphics[width=0.4\linewidth]{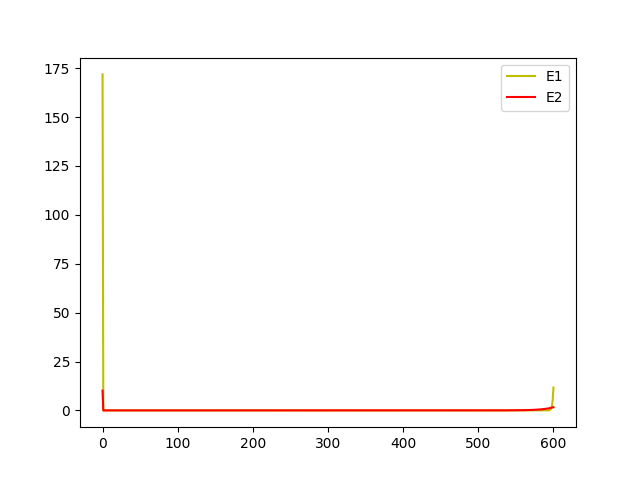}}\qquad\qquad\\
		\subfloat[Infected $I_{1}$ and $I_{2}$.\label{fig:I1I2-EDO}]{\includegraphics[width=0.4\linewidth]{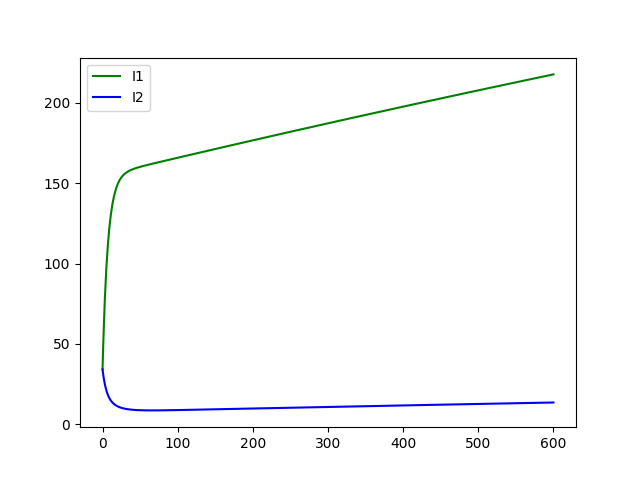}}\qquad\qquad
		\subfloat[Infected $I_{1}$ and $I_{2}$.\label{fig:I1I2-Optcont}]{\includegraphics[width=0.4\linewidth]{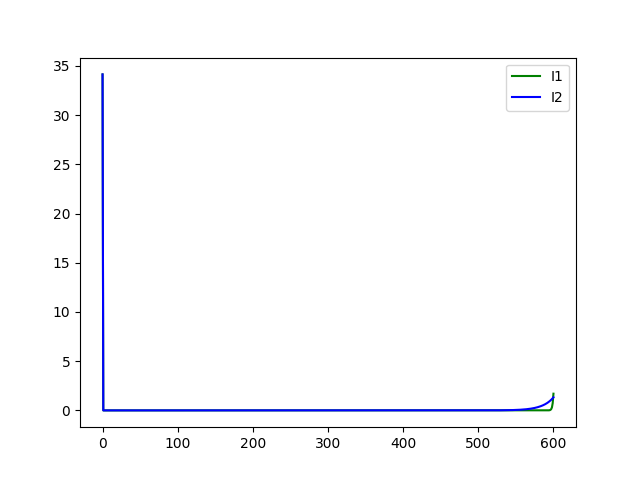}}\qquad\qquad\\
		\caption{Plot of states with and without controls of test 1. Without controls on the left and with controls on the right.}
		\label{fig:EDO-Optcont}		
	\end{figure}
		\begin{figure}[H]
			\centering
			\subfloat[Susceptible $S$ and recovered $R$ cases.\label{fig:SR-EDO-2-sc}]{\includegraphics[width=0.4\linewidth]{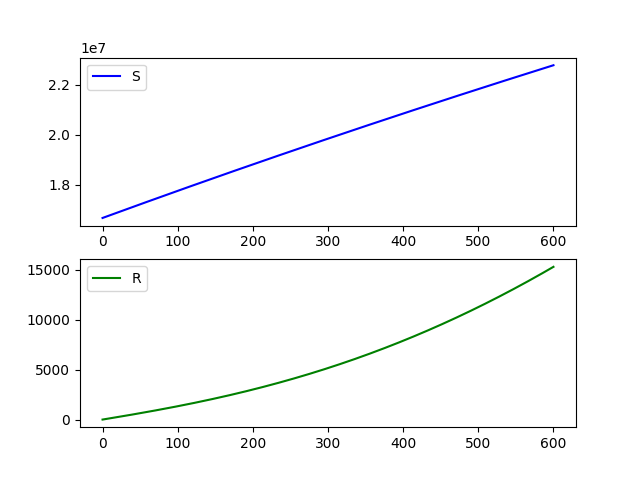}}\qquad\qquad 
			\subfloat[Susceptible $S$ and recovered $R$ cases.\label{fig:SR-Optcont-2-sc}]{\includegraphics[width=0.4\linewidth]{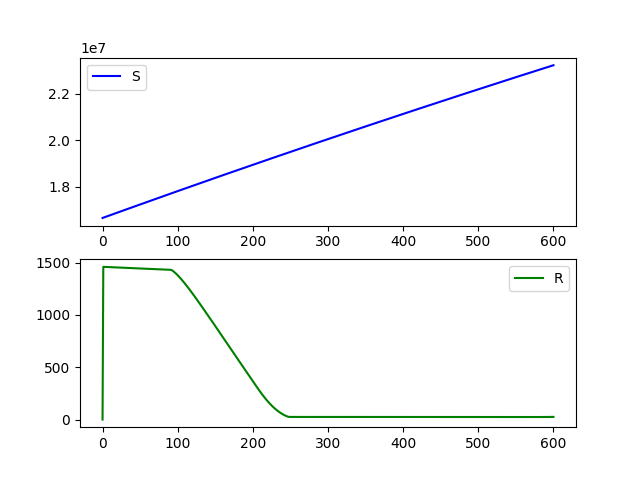}}\qquad\qquad \\
			\quad\subfloat[Exposed $E_{1}$ and $E_{2}$.\label{fig:E1E2-EDO-2-sc}]{\includegraphics[width=0.4\linewidth]{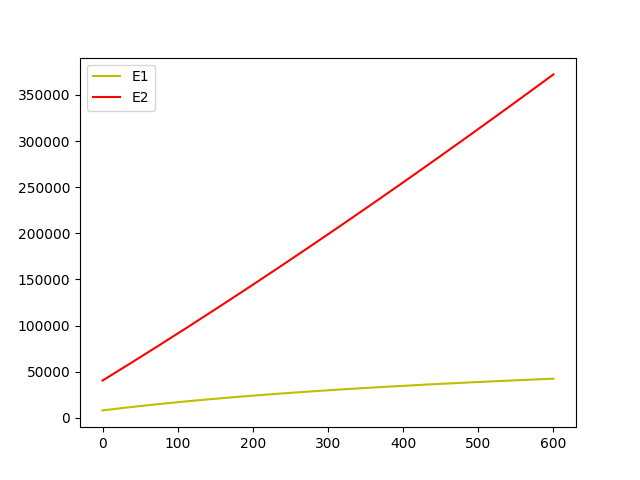}}\qquad\qquad
			\quad\subfloat[Exposed $E_{1}$ and $E_{2}$.\label{fig:E1E2-Optcont-2-sc}]{\includegraphics[width=0.4\linewidth]{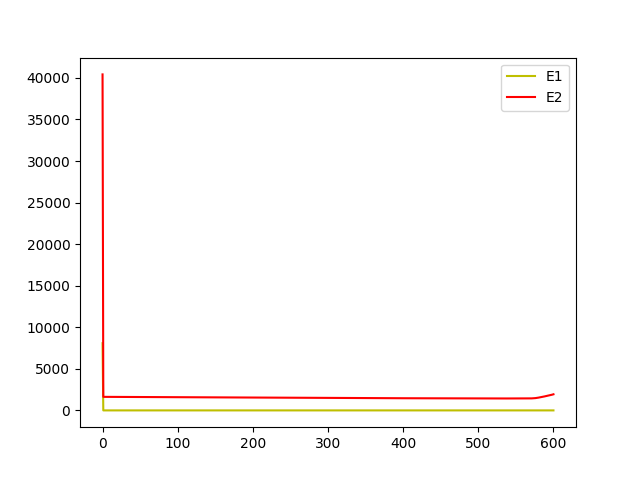}}\qquad\qquad\\
			\subfloat[Infected $I_{1}$ and $I_{2}$.\label{fig:I1I2-EDO-2-sc}]{\includegraphics[width=0.4\linewidth]{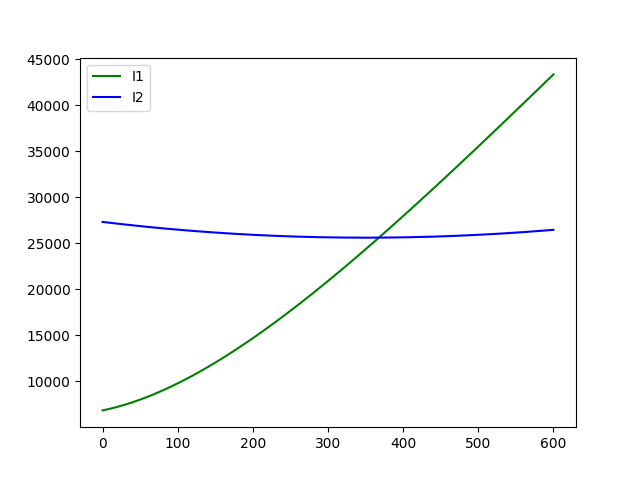}}\qquad\qquad
			\subfloat[Infected $I_{1}$ and $I_{2}$.\label{fig:I1I2-Optcont-2-sc}]{\includegraphics[width=0.4\linewidth]{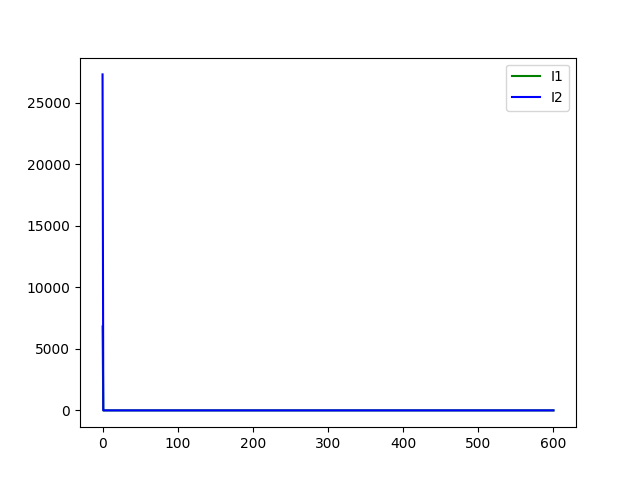}}\qquad\qquad\\
			\caption{Plot of states with and without controls of test 2. Without controls on the left and with controls on the right.}
			\label{fig:EDO-Optcont-2-sc}		
		\end{figure}
		\begin{figure}[H]
			\centering
			\subfloat[Susceptible $S$.\label{fig:CompS}]{\includegraphics[width=0.4\linewidth]{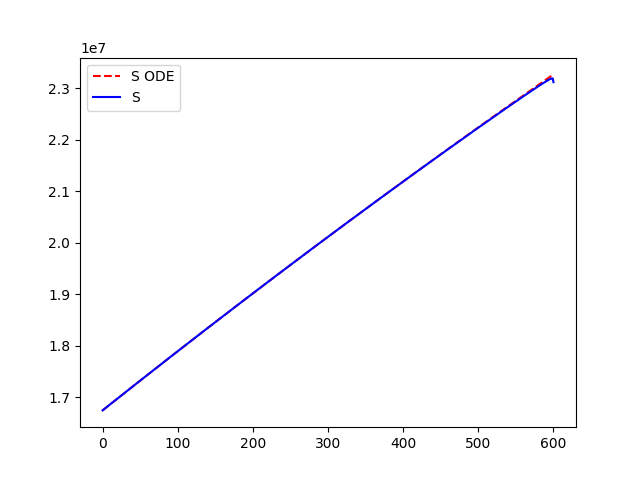}}\qquad\qquad 
			\subfloat[Recovered $R$.\label{fig:CompR}]{\includegraphics[width=0.4\linewidth]{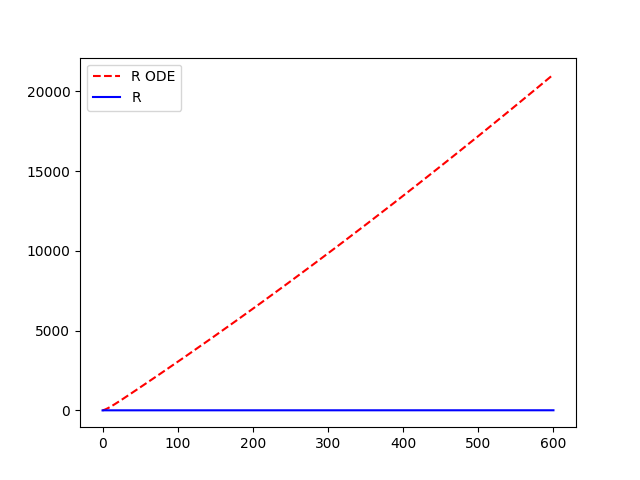}}\qquad\qquad \\
			\quad\subfloat[Exposed $E_{1}$.\label{fig:CompE1}]{\includegraphics[width=0.4\linewidth]{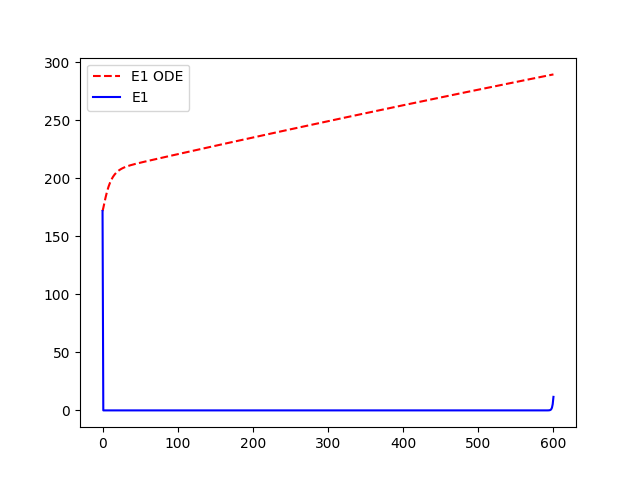}}\qquad\qquad
			\subfloat[Exposed $E_{2}$.\label{fig:CompE2}]{\includegraphics[width=0.4\linewidth]{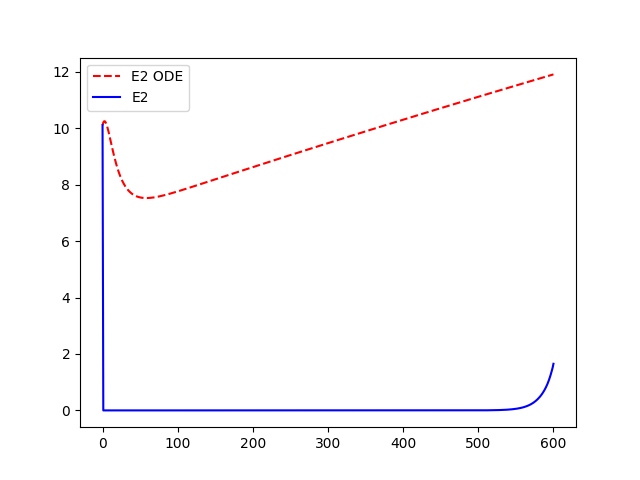}}\qquad\qquad\\
			\quad\quad\subfloat[Infected $I_{1}$.\label{fig:CompI1}]{\includegraphics[width=0.4\linewidth]{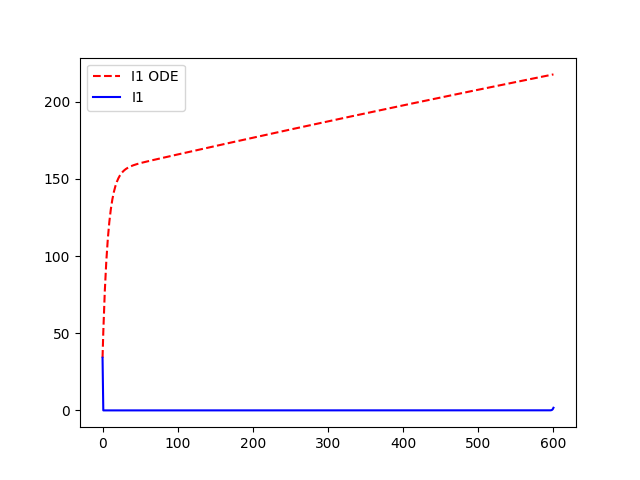}}\qquad\qquad
			\subfloat[Infected $I_{2}$.\label{fig:CompI2}]{\includegraphics[width=0.4\linewidth]{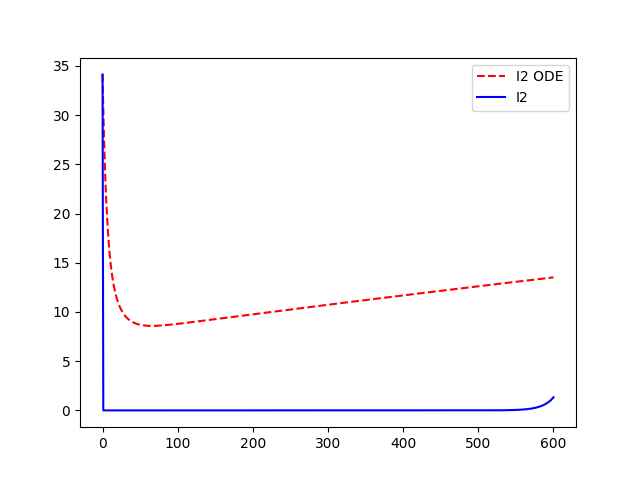}}\qquad\qquad
			\caption{Comparative plot of states with and without controls of test 1. Without controls in red dotted lines and with controls in blue lines.}
			\label{fig:Comparative}		
		\end{figure}
			\begin{figure}[H]
				\centering
				\subfloat[Susceptible $S$.\label{fig:CompS-2-sc}]{\includegraphics[width=0.4\linewidth]{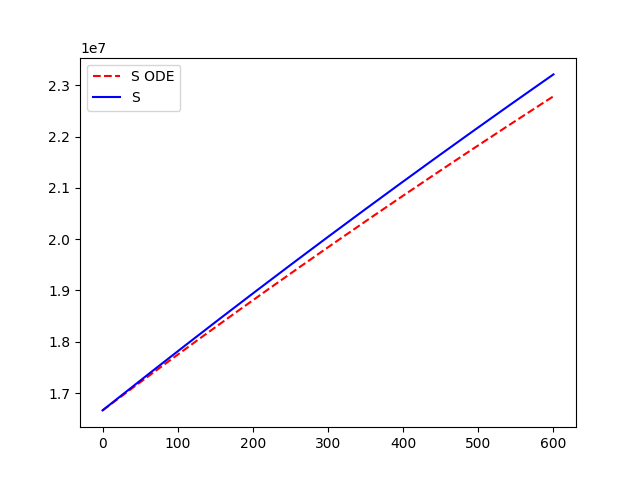}}\qquad\qquad 
				\subfloat[Recovered $R$.\label{fig:CompR-2-sc}]{\includegraphics[width=0.4\linewidth]{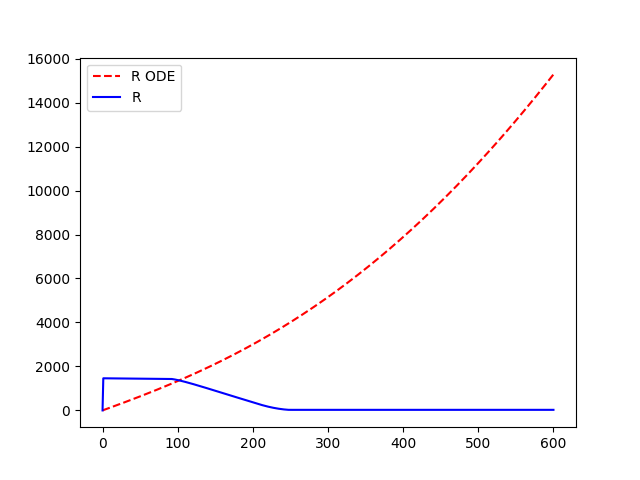}}\qquad\qquad 
				\quad\subfloat[Exposed $E_{1}$\label{fig:CompE1-2-sc}]{\includegraphics[width=0.4\linewidth]{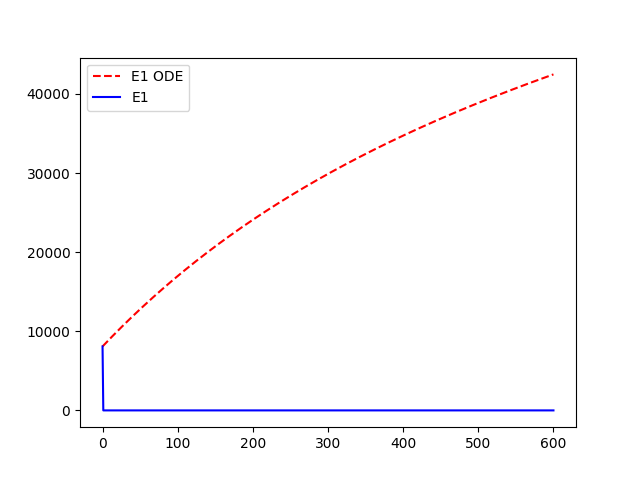}}\qquad\qquad
				\subfloat[Exposed $E_{2}$.\label{fig:CompE2-2-sc}]{\includegraphics[width=0.4\linewidth]{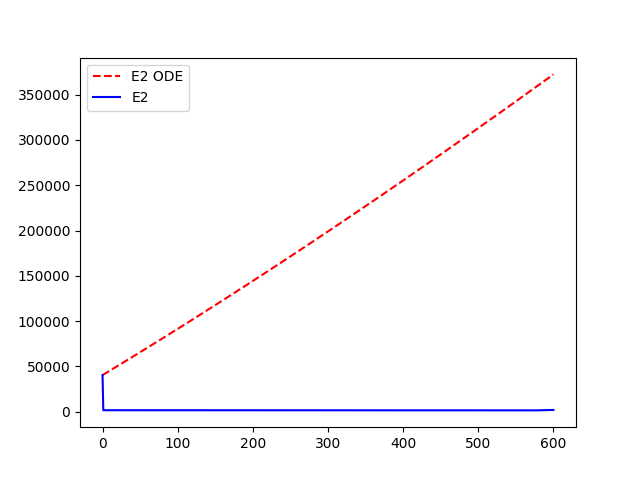}}\qquad\qquad\\
				\quad\subfloat[Infected $I_{1}$.\label{fig:CompI1-2-sc}]{\includegraphics[width=0.4\linewidth]{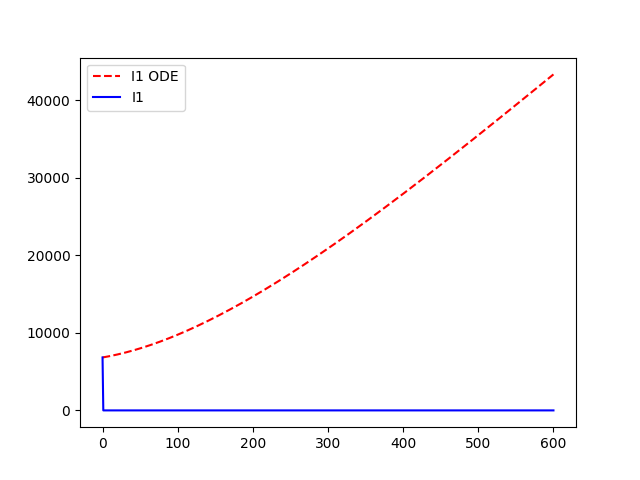}}\qquad\qquad
				\subfloat[Infected $I_{2}$.\label{fig:CompI2-2-sc}]{\includegraphics[width=0.4\linewidth]{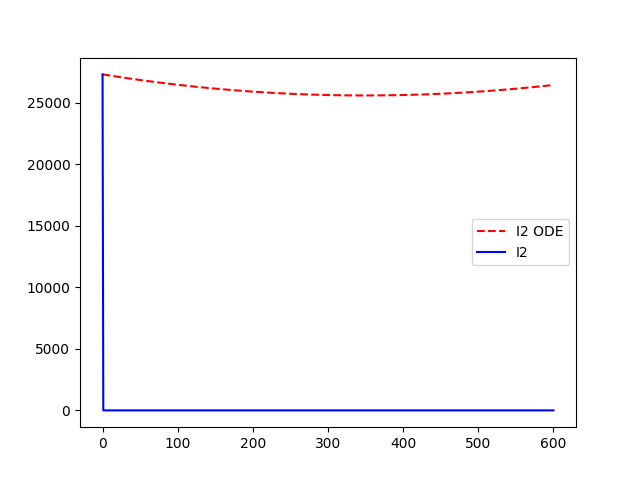}}\qquad\qquad
				\caption{Comparative plot of states with and without controls of test 2. Without controls in red dotted lines and with controls in blue lines.}
				\label{fig:Comparative-2-sc}		
			\end{figure}
			\noindent	Figures \ref{fig:EDO-Optcont}, \ref{fig:EDO-Optcont-2-sc}, \ref{fig:Comparative} and \ref{fig:Comparative-2-sc} show the dynamics of two-strain infection model with and without controls. We clearly observe that with control, the amount of susceptible is higher than without control. Hence, the controls have played an essential role in maximizing the susceptible individuals. Moreover,  in both strains, infected and exposed individuals are reduced with control, while without control, they remain at a strictly positive level. The convergence towards the free endemic equilibrium with control is observed. More precisely, all the states converge toward the endemic-free equilibrium $(\ \frac {\Lambda}{\delta},0,0,0,0,0)$, which means that the infection can be eradicated. 		
			\begin{figure}[H]
				\centering
				\subfloat[Test 1\label{fig:Cum-Optcont}]{\includegraphics[width=0.4\linewidth]{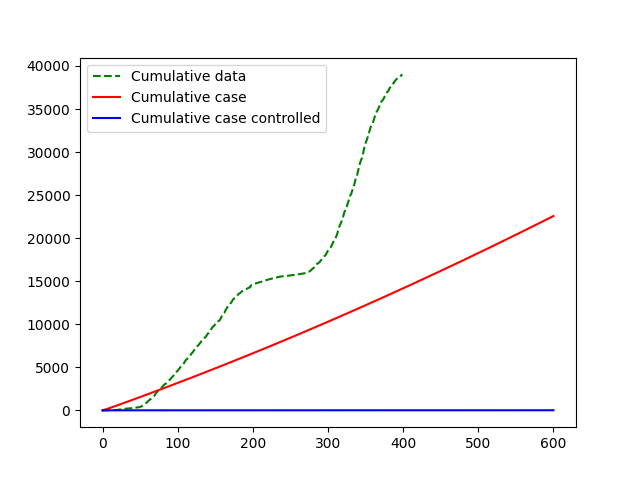}}\qquad\qquad 
				\subfloat[Test 2\label{fig:Cum-Optcont-2-sc}]{\includegraphics[width=0.4\linewidth]{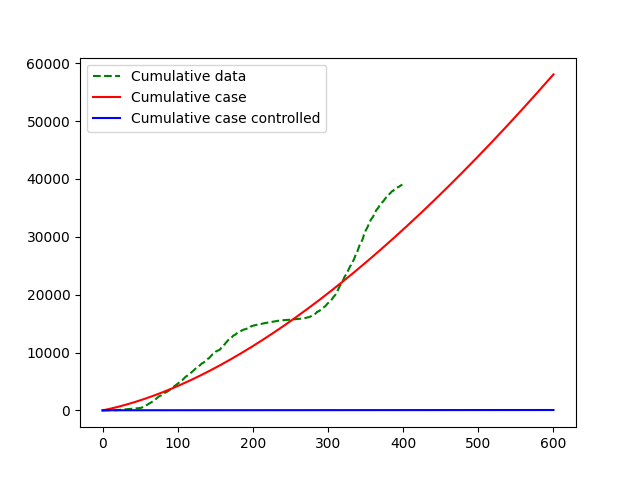}}\qquad\qquad
				
				\caption{Plot of cumulative case. The COVID-19 cumulative case data is in green, the cumulative case without control is in red, and the minimal cumulative case is in blue.}
				\label{fig:cumcase}	
			\end{figure}
			\noindent The figure \ref{fig:cumcase} shows a comparison of cumulative case data, the cumulative case with and without controls.
			
			\begin{figure}[H]
				\centering
				\subfloat[\label{fig:u1-Optcont}]{\includegraphics[width=0.4\linewidth]{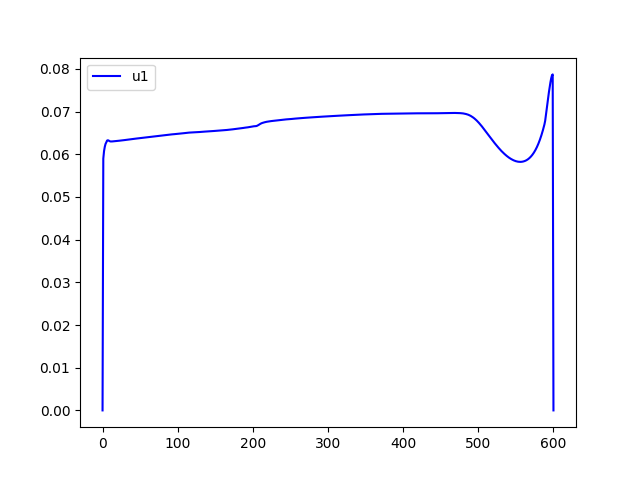}}\qquad\qquad 
				\subfloat[\label{fig:u2-Optcont}]{\includegraphics[width=0.4\linewidth]{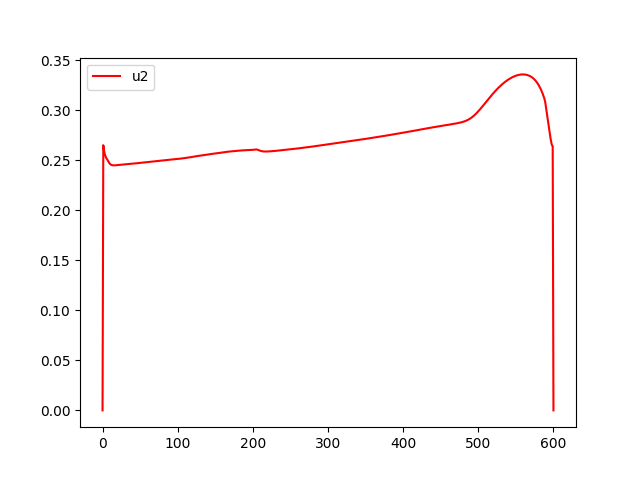}}\qquad\qquad
	
				\caption{Plot of controls of the test 1.}
				\label{fig:controls-Optcont}	
			\end{figure}
			
			\begin{figure}[H]
				\centering
				\subfloat[\label{fig:u1-Optcont-2}]{\includegraphics[width=0.4\linewidth]{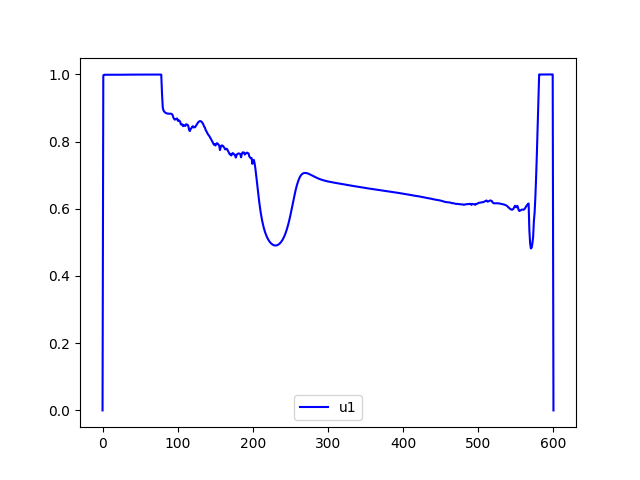}}\qquad\qquad 
				\subfloat[\label{fig:u2-Optcont-2}]{\includegraphics[width=0.4\linewidth]{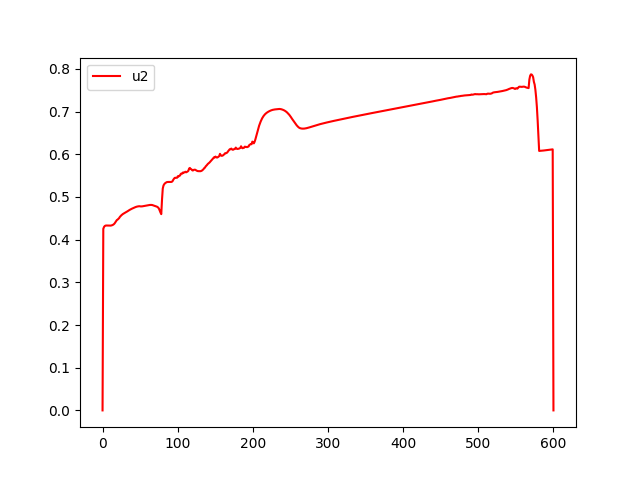}}\qquad\qquad
	
				\caption{Plot of controls of the test 2.}
				\label{fig:controls-Optcont-2}	
			\end{figure}
			\noindent In both tests the controls $u_{1}$ and $u_{2}$ are in high level as shown in the figures \ref{fig:controls-Optcont} and \ref{fig:controls-Optcont-2}.
	
			\section*{Estimation of parameters}
			\label{estima}
\begin{figure}[H]
				\centering
				\subfloat[Cumulative case data in red and fit function $TNI(t)$ in blue.  \label{fitcumul}]{\includegraphics[width=0.4\linewidth]{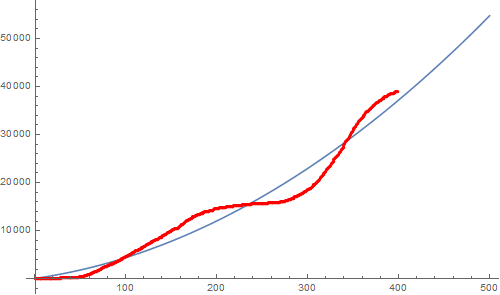}}\qquad\qquad 
				\subfloat[Recovered case data in red and fit function $TR(t)$ in blue.\label{firecov}]{\includegraphics[width=0.4\linewidth]{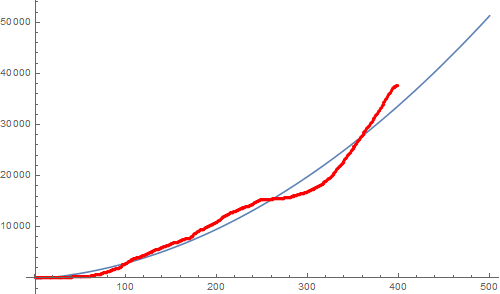}}
				\caption{Fit of cumulative cases and recovered data from 2020, March 02 to 2021 April 04.}
				\label{fitdata}	
			\end{figure}

			The estimation of the parameters of the model \eqref{prob} is done by using techniques in \cite{Balde2020},  \cite{Balde22020} and \cite{Mansaletal}. We fit respectively the cumulative and recovered case data of $COVID-19$ in Senegal, from 2020 March 02 to 2021 April 04, with the two polynomial functions $TNI(t)=a t^2+ bt+c$ and $TR(t)=e t^2+ ft+g$, see figure \ref{fitdata}. In addition, we assume that these  functions can be given in integral form as $TNI(t)=\displaystyle\int_{0}^{t}\mu_{1}E_{1}(s)+\mu_{2}E_{2}(s)ds$ and $TR(t)=\displaystyle\int_{0}^{t}\gamma_{1}I_{1}(s)+\gamma_{2}I_{2}(s)ds$.\\
			Derivating $TNI$ gives 
			\begin{align*}
				\mu_{1}E_{1}(t)+\mu_{2}E_{2}(t)=2 a t +b&\\
				\mu_{1}E_{1,0}+\mu_{2}E_{2,0}=b&\\
			\end{align*}
			We suppose that $\mu_{2}E_{2,0}=\xi_{1}\mu_{1}E_{1,0}$ implying $(1+\xi_{1})\mu_{1}E_{1,0}=b$. Then, 
			\begin{align*}
				\mu_{1}=\displaystyle \frac{b}{(1+\xi_{1})E_{1,0}}
			\end{align*}
			Derivating again, we obtain 
			\begin{align*}
				\mu_{1}\dot{E}_{1}(t)+\mu_{2}\dot{E}_{2}(t)=2 a&\\
				\mu_{1}\dot{E}_{1,0}+\mu_{2}\dot{E}_{2,0}=2a&\\
			\end{align*}
			Supposing that $\mu_{2}\dot{E}_{2,0}=\xi_{1}\mu_{1}\dot{E}_{1,0}$, we obtain $\mu_{1}\dot{E}_{1,0}=\displaystyle\frac{2a}{1+\xi_{1}}$. We get 
			\begin{align*}
				E_{1,0}=\frac{b}{2a}\dot{E}_{1,0}	
			\end{align*}
			Some calculations give 
			\begin{align*}
				\mu2 =\displaystyle \xi_{1}\mu_{1}\frac{\dot{E}_{1,0}}{\dot{E}_{2,0}},\quad E_{2,0} =\displaystyle E_{1,0}\frac{\dot{E}_{2,0}}{\dot{E}_{1,0}}.&
			\end{align*}
			Derivating $TR(t)$ gives 
			\begin{align*}
				\gamma_{1}I_{1}(t)+\gamma_{2}I_{2}(t)=2 e t +f&\\
				\gamma_{1}I_{1,0}+\gamma_{2}I_{2,0}=f&\\
			\end{align*}
			We suppose that $\gamma_{2}I_{2,0}=\xi_{2}\gamma_{1}I_{1,0}$. Doing the same work as above, we obtain
			\begin{align*}
				\gamma_{1}=\displaystyle \frac{b}{(1+\xi_{2})I_{1,0}},\quad \gamma_2 =\displaystyle \xi_{2}\gamma_{1}\frac{\dot{I}_{1,0}}{\dot{I}_{2,0}}, \quad I_{1,0} =\displaystyle f\frac{\dot{I}_{1,0}}{2e}, \quad I_{2,0} =\displaystyle I_{1,0}\frac{\dot{I}_{2,0}}{\dot{I}_{1,0}}&  
			\end{align*}
			Using the second equation of the model \ref{prob} and setting $u_{1}=u_{2}=0$, we obtain \\
			\begin{align*}
				\displaystyle \dot{E}_{1} = \displaystyle\frac{\beta_1 SI_1}{1+k_1 I_1}-(\mu_1+\delta)E_1&\\
				\displaystyle \dot{E}_{1,0} = \displaystyle\frac{\beta_1  S_{0}I_{1,0}}{1+k_{1} I_{1,0}}-(\mu_1+\delta)E_{1,0}&\\
				\frac{2a}{\mu_{1}(1+\xi_{1})}=\displaystyle\frac{\beta_1  S_{0}I_{1,0}}{1+k_{1} I_{1,0}}-(\mu_1+\delta)\frac{b}{\mu_{1}(1+\xi_{1})}&\\
				\beta_{1} = \displaystyle \frac{(1+k_{1} I_{1,0})(2a+b\mu_{1}+\delta b)}{\mu_{1}(1+\xi_{1})S_{0}I_{1,0}}
			\end{align*}
			Using again the third equation of the model \ref{prob} and doing the same work as above, give 
			\begin{align*}
				\beta_{2} = \displaystyle \frac{\xi_{1}(1+k_{2} I_{2,0})(2a+b\mu_{1}+\delta b)}{\mu_{2}(1+\xi_{1})S_{0}I_{2,0}}
			\end{align*}
			
			\noindent We use $32.9\% $ of year $2018$ for the birth rate from \cite{wiki}. 
	
			Then, the recruitment is $\Lambda=32.9\% T/365$ by day. The death rate is $7.9\% $ by year at $2018$.   
			\noindent Now, we discusse about the coefficients $C_{1},\ C_{2}$. We simulate two tests. For both tests, the values of the coefficients are related to the infection rate. We consider that if the infection rate is such that $\beta_{1}<\beta_{2}$, then the propagation of strain 2 is more than that of strain 1. The more the infection spreads through the population, the greater the treatment expenses. Then, we set $C_{1}=100/\beta_{2}$ and $C_{2}=100/\beta_{1}$. 			

			\section{Conclusion}
			In this work, we have studied numerically and theoretically a mathematical model describing the dynamics of a two-strain SEIR epidemic model. The model contains a system of nonlinear differential equations describing the interaction between the susceptible, the first, and the second strain exposed individuals,  the first and the second strain infected individuals, and the recovered ones. Two saturated rates and two treatments were incorporated into the model. The well-posedness of the model was established in terms of positivity and boundedness of solutions. The basic reproduction number was calculated as a function of the reproduction numbers of the first and second strains. The global stability of the disease-free equilibrium was fulfilled. We have performed an estimation study of the problem parameters. In addition,  the model numerical simulation was compared with COVID-19 clinical data. The optimal control study was achieved by using the Pontryagin minimum principle. Numerical simulations have shown the importance of therapy in minimizing the infection's effect. It was shown that the disease severity decreased remarkably when good therapies were administrated.
			
			\section*{Acknowledgements}
			\textit{Karam Allali is grateful to the support from the Laboratory Mathematics, Computer Science and Applications, University Hassan II of Casablanca, and the hospitality of Babacar Mbaye Ndiaye at LMDAN in the University Cheikh Anta Diop of Dakar during the research collaboration visit}.

			\bibliographystyle{plain}
			\bibliography{twostrainbib}

\end{document}